\documentclass[%
 aip,
 jmp,%
 amsmath,amssymb,
 reprint,%
]{revtex4-2}

\usepackage{graphicx}
\usepackage{dcolumn}
\usepackage{bm}

\begin{document}

\preprint{AIP/123-QED}

\title{Holographic hyperbranched polymer nanocomposite grating with exceptionally large neutron scattering length density modulation amplitudes}

\author{Elhoucine Hadden}
\email{elhoucine.hadden@physics.uu.se}
 \affiliation{Faculty of Physics \& Vienna Doctoral School in Physics, University of Vienna, 1090 Vienna, Austria}
 \altaffiliation{Current affiliation: Department of Physics and Astronomy, Uppsala University, Box 516, 75120 Uppsala, Sweden}
\affiliation{Institut Laue-Langevin, 71 avenue des Martyrs, CS 20156, 38042 Grenoble Cedex 9, France}

\author{J\"{u}rgen Klepp} 
\email{juergen.klepp@univie.ac.at}
\affiliation{Faculty of Physics, University of Vienna, 1090 Vienna, Austria} 

\author{Martin Fally}
\affiliation{Faculty of Physics, University of Vienna, 1090 Vienna, Austria}

\author{Tobias Jenke} 
\affiliation{Institut Laue-Langevin, 71 avenue des Martyrs, CS 20156, 38042 Grenoble Cedex 9, France}

\author{Joachim Kohlbrecher} 
\affiliation{PSI Center for Neutron and Muon Sciences, 5232 Villigen PSI, Switzerland} 

\author{Tomoko Shimada} 
\affiliation{Department of Engineering Science, University of Electro-Communications, 1-5-1 Chofugaoka, Chofu, Tokyo 182-8585, Japan}

\author{Asako Narita} 
\affiliation{Department of Engineering Science, University of Electro-Communications, 1-5-1 Chofugaoka, Chofu, Tokyo 182-8585, Japan}

\author{Juro Oshima} 
\affiliation{Materials Research Laboratories, Nissan Chemical Corporation, 488-6 Suzumi,Funabashi, Chiba 274-0052, Japan}

\author{Yasuo Tomita} 
\affiliation{Department of Engineering Science, University of Electro-Communications, 1-5-1 Chofugaoka, Chofu, Tokyo 182-8585, Japan}

\date{\today}

\begin{abstract}
Nanoparticle–polymer composite gratings incorporating ultrahigh-refractive-index hyperbranched polymers as organic nanoparticles have demonstrated exceptional light optical properties, yet their potential for neutron diffraction applications remains unexplored.
We report on the neutron optical properties of a holographically structured hyperbranched-polymer–dispersed nanocomposite grating at a quasi-monochromatic neutron wavelength of 2\,nm.
We show that neutron diffraction measurements performed at the SANS-I instrument of the Paul Scherrer Institute (Switzerland) reveal exceptionally high neutron scattering length density modulation amplitudes.
These scattering length density modulation amplitudes are the highest reported to date.
Very high neutron diffraction efficiency is expected with the use of thicker uniform gratings and longer neutron wavelengths, with low angular and wavelength selectivity constraints.
\end{abstract}

\maketitle


Neutron interferometry is a uniquely sensitive technique for probing both material properties and fundamental quantum phenomena \cite{2015_RauchWerner_Book}. Its ability to detect coherent phase shifts enables precision measurements of scattering length densities (SLDs) \cite{2020_Haun_PRL} and has supported landmark experiments, including the demonstration of \( 4\pi \)-symmetry of fermionic wavefunctions, gravitational quantum interference, and neutron spin-path entanglement \cite{2024_Danner_Nat}. Extending neutron interferometry into the cold neutron (CN) and very cold neutron (VCN) wavelength regimes enhances sensitivity to quantum effects.
A first generation of VCN interferometers has been operated since the end of the 1980s, notably with the establishment of a dedicated platform at the PF2/VCN instrument at the Institut Laue Langevin (ILL) \cite{1989_Eder_NIMA}.
Despite the scientific interest in such experiments, the boom has gradually declined due to two main limitations: the inherently low flux of VCN beams and the lack of efficient neutron-optical components adapted to longer neutron wavelengths. 
Addressing the need for wavelength-adapted optical elements is crucial to enable the next generation of highly sensitive slow neutron interferometers.

Regarding neutron source limitations, the development of advanced neutron facilities is underway, incorporating dedicated VCN sources to overcome flux constraints \cite{2010_Mezei_JNR}. For instance, the HighNESS project aims to establish a high-flux VCN source at the European Spallation Source (ESS), promising substantial improvements in available neutron flux \cite{2024_Santoro_NSE}. However, fully exploiting these advances requires complementary progress in neutron optical components. Traditional neutron optics for interferometry such as perfect single-crystals suffer from intrinsic limitations at VCN wavelengths. 
An alternative approach is the use of holographic optical elements (HOEs) by transforming blends of photosensitive materials into artificial periodic structures. 
Several material classes along this line have been reported so far.
These include gratings recorded in thick binder-based PMMA photopolymer \cite{1990_Rupp_PRL}, holographic polymer-dispersed liquid crystals (HPDLCs) \cite{2006_Fally_PRL}, and nanoparticle polymer composites (NPCs) \cite{2016_Tomita_JMO, 2010_Fally_PRL_NPCs}.
It was found that the 1\,mm-order thickness of the PMMA gratings resulted in low neutron diffraction efficiency due to the Pendellösung interference effect \cite{2016_Tomita_JMO}. 
HPDLCs were also found to suffer from high anisotropy causing significant light scattering during recording and thereby resulting in low neutron diffraction efficiency.
On the other hand, NPCs, which were originally developed for holographic gratings with high diffraction efficiency at visible wavelengths, low polymerization shrinkage, and high thermal stability—desirable for holographic data storage, nonlinear optics, and wearable displays \cite{2016_Tomita_JMO}—have also been found efficient at slow neutron wavelengths.
It was also found recently that the use of nanodiamonds, with their large SLD value, as nanoparticles in NPCs provided high performance in neutron diffraction \cite{2020_Tomita_PRA, 2022_Hadden_SPIE, 2024_Hadden_APL}.

In this work, we describe an experimental investigation of neutron diffraction properties of a holographic NPC grating incorporating hyperbranched polymer (HBP) as organic nanoparticles—originally developed for other photonic applications \cite{2016_Tomita_OL_HBPs, 2020_Tomita_OE_HBP, 2021_Narita_OME}—at a mean neutron wavelength of 2\,nm.

\begin{figure}[htbp]
\begin{center}
 \includegraphics[clip, width=80mm]{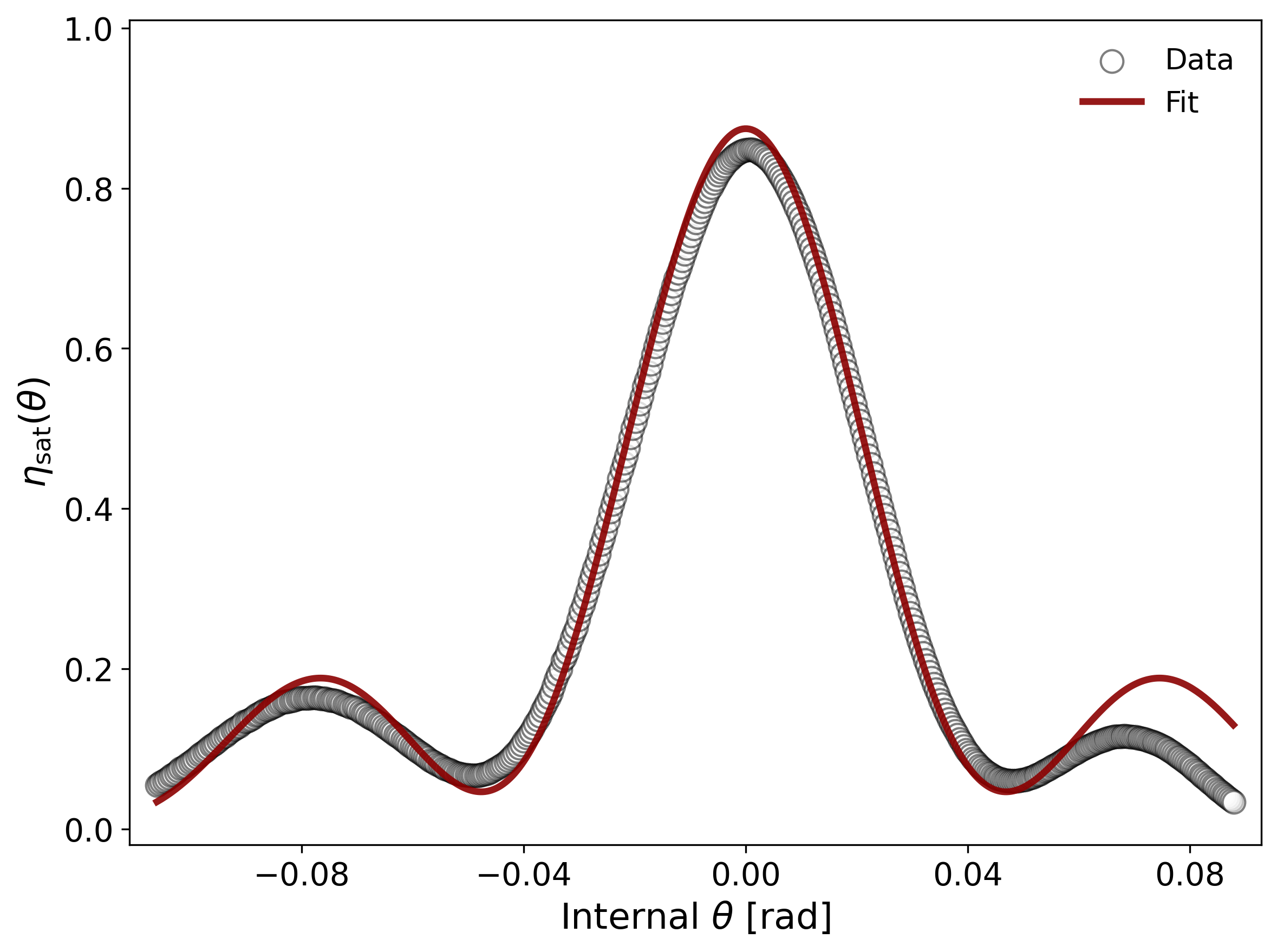}
\end{center}
\caption{Bragg-angle detuning curve of $\eta_\text{\tiny{sat}}$ probed at 532\,nm (empty circles), together with the model fit from standard least-squares minimization (solid line).}
\label{fig:f1}
\end{figure}

\indent The sample preparation and recording procedures follow those described in Ref.~\citenum{2021_Narita_OME}.
Holographic recording of a 500\,nm-period grating was performed in a red-sensitive, HBP-dispersed NPC film. The process employed a conventional two-wave mixing setup and a single-longitudinal mode laser operating at 640\,nm.
The angular selectivity curve of the saturated diffraction efficiency $\eta_\text{\tiny{sat}}(\theta)$, where $\theta$ is the Bragg-angle detuning-was measured by a probe beam at 532\,nm as shown in Figure~\ref{fig:f1}. 

\noindent The observed asymmetry in the data is attributed to secondary scattering, the origin of which will be explained in the subsequent analysis. In addition, the lifted side-lobe minima indicate a grating decay along the thickness direction \( z \) \cite{2016_Tomita_OL_HBPs, 2020_Tomita_OE_HBP}. We performed a least-squares fit to \( \eta_\text{\tiny{sat}}(\theta) \) using Uchida's formula \cite{1973_Uchida_JOSA}, based on the beta-value method (BVM) \cite{1992_Sheridan_JMO}. The model includes the thickness-dependent first-order refractive index modulation amplitude \( n_1(z) = n_{10}\exp(-z/L) \), where \( d \) is the physical film thickness and \( L \) is the effective decay length of the recorded grating.
The best-fit parameters are a grating thickness of \( d \approx 8.4\,\mu\mathrm{m} \), a modulation amplitude of \( n_{10} = (5.6 \pm 0.0) \times 10^{-2} \), and \( L \approx 9.3\,\mu\mathrm{m} \).
\( L \) is comparable to \( d \), attributed to grating attenuation from holographic scattering during recording \cite{2007_Suzuki, 2020_Tomita_OE_HBP} in a sample with extremely large \( n_{10} \).
The effective grating thickness \( d_{\textit{eff}} \), defined as \( L \big[1 - \exp(-d/L) \big] \), is \( 5.5\,\mu\mathrm{m} \), and the thickness-averaged modulation amplitude \( \langle n_1(z) \rangle \), defined as \( n_{10} \, d_{\textit{eff}} / d \), is \( (3.7 \pm 0.2) \times 10^{-2} \).
The corresponding modulation of the nanoparticle volume fraction is calculated via \( a_1 \Delta f = \langle n_1 \rangle / |\text{n}_\text{\tiny{NP}} - \text{n}_\text{\tiny{P}}| \) \cite{2020_Tomita_PRA}, yielding \( a_1 \Delta f = 0.114 \) for \( n_{\text{NP}} = 1.82 \) and \( n_{\text{P}} = 1.50 \) \cite{2016_Tomita_OL_HBPs}.
This value is roughly 8~times higher than the best previously reported for nanodiamond-based NPC gratings \cite{2024_Hadden_APL, 2022_Hadden_SPIE}, which motivates the subsequent investigation of its neutron-optical performance.

\begin{figure}[htbp]
\begin{center}
 \includegraphics[clip, width=80mm]{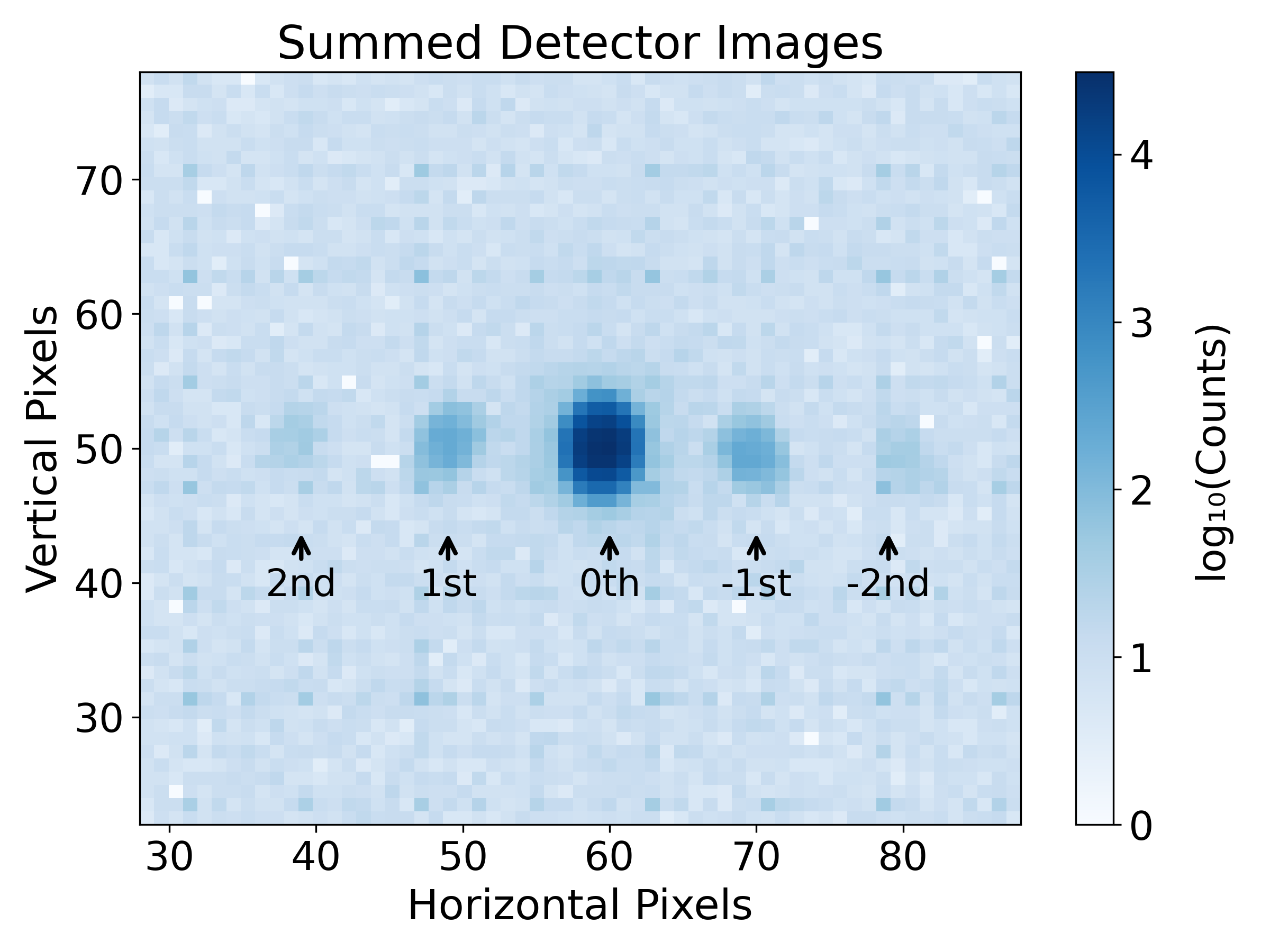}
\end{center}
\caption{Accumulated detector image obtained by summing the recorded counts at each pixel over all angular positions of the $\theta$-scan, displayed on a logarithmic color scale of the total counts.}
\label{fig:f3}
\end{figure}

\indent Neutron diffraction measurements were performed at the SANS-I instrument at the Swiss Spallation Neutron Source (SINQ), Paul Scherrer Institute (PSI). 
A mean neutron wavelength of $\lambda=2$\,nm was selected using a helical slot velocity selector, with a relative wavelength spread of $\Delta\lambda / \lambda \approx 0.1$.
Beam collimation was achieved using two apertures separated by 18\,m: a 30-mm aperture at the entrance and a 4-mm aperture behind the sample. The resulting beam divergence of about 1.9\,mrad is much narrower than the expected diffraction peak width, approximated by $2\Lambda/d \approx 0.12$\,rad, ensuring sufficiently high angular resolution.
To enable angular $\theta$-scans around normal incidence, the sample was mounted on a rotational stage. It was tilted at a fixed angle $\zeta = 45^\circ$ around the grating vector of the unslanted structure in order to increase the effective thickness for diffraction (see, for instance, Ref.~\citenum{2010_Fally_PRL_NPCs}).
A two-dimensional $^3$He neutron detector with $128\times128$ pixels of $7.5\times7.5$\,mm$^2$ size is placed 18\,m away from the sample and was used to record the diffracted intensities.
The data collected during the $\theta$-scan were summed over all angular positions to construct the accumulated detector image shown in Figure~\ref{fig:f3}.

\begin{figure}[htbp]
\begin{center}
 \includegraphics[clip, width=80mm]{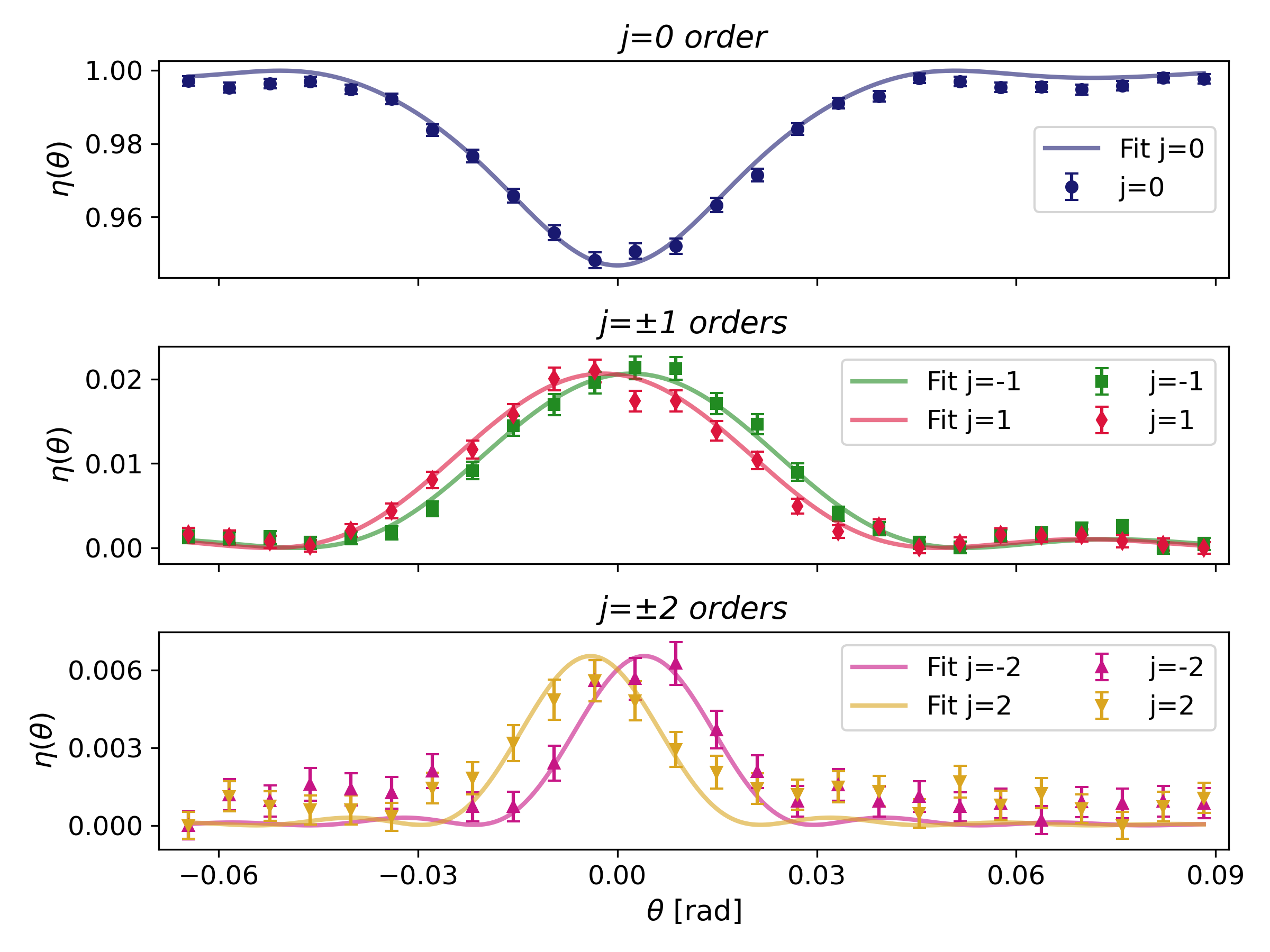}
\end{center}
\caption{Neutron diffraction efficiency curves of the HBP-dispersed NPC grating (symbols) and corresponding fits (solid lines) at a neutron wavelength of 2\,nm.}
\label{fig:f4}
\end{figure}

\noindent Diffraction spots up to the second order are clearly visible around the intense forward-diffracted (0th order) beam. A logarithmic scale is used to enhance the visibility of the weaker second-order peaks. Further experimental details and the complete data reduction procedure are provided in Ref.~\citenum{2024_Hadden_thesis}. The derived diffraction efficiency curves along with corresponding fits are presented in Figure~\ref{fig:f4}.

\noindent The diffraction efficiency curves were analyzed using a truncated first-order seven-coupled waves (7-CW) analysis with BVM boundary conditions~\cite{2024_Hadden_thesis, 2016_Klepp_JPCS}.
The model accounts for the first- and second-order Fourier components of the SLD modulation. 
This approach is justified by the presence of significant diffraction into five orders ($j=0$, $\pm1$, $\pm2$).
To account for all coupling terms up to the second diffraction orders, a 7-CW analysis is required. This is accomplished at reasonable computational costs.
In the neutron diffraction analysis, the $j$-th SLD modulation amplitude \( \Delta(b_c \rho)_j \) (\( j = 1, 2 \)) is related to the corresponding neutron refractive index modulation amplitude \( n_j \) by \( \Delta(b_c \rho)_j = \frac{2\pi}{\lambda^2} n_j \), with \( |\Delta(b_c \rho)_j| = a_j \Delta f \, \left| (b_c \rho)_{\mathrm{NP}} - (b_c \rho)_{\mathrm{P}} \right| \) \cite{2020_Tomita_PRA}.
The fit yields an extremely large first-order thickness-averaged SLD modulation amplitude, $\langle \Delta (b_c \rho)_1 \rangle = (14.9 \pm 0.2)$\,$\mu$m$^{-2}$. This marks a net improvement over the highest value recently reported for nanodiamond-dispersed NPC gratings~\cite{2024_Hadden_APL}.
A comparable magnitude to the latter is found for the second-order component, $\langle \Delta (b_c \rho)_2 \rangle = (-8.2 \pm 0.3)$\,$\mu$m$^{-2}$. 
The negative sign reflects a relative phase $\varphi_2 = \pi$ between the first- and second-order components of the HBP-dispersed NPC grating.
These results demonstrate that, even without specific optimization, HBP-based NPC gratings exhibit outstanding neutron-optical performance, setting a new benchmark for neutron SLD modulation amplitudes. This is primarily attributed to the large volume fraction modulation extracted from light-optical analysis, which exceeds that of previously reported nanodiamond-based gratings by nearly an order of magnitude. 
Despite the modest intrinsic SLD contrast between the HBP nanoparticles and the host polymer compared to systems incorporating inorganic nanoparticles, the high photosensitivity and favorable diffusion properties promote efficient phase separation during holographic recording. This enables SLD modulation amplitudes that exceed conventional limits.

\indent In summary, HBP-based NPC gratings represent promising candidates for neutron-optical applications in the cold and very cold neutron regimes.
The material exhibits exceptionally high neutron SLD modulation amplitudes, surpassing previous benchmarks.
While the current grating thickness limits diffraction efficiency at $\lambda=2$\,nm, a moderate thickness increase is expected to substantially enhance the performance.
Future research should address the non-trivial challenge of optimizing material composition and recording conditions to allow moderate increases in grating thickness while maintaining large modulation contrast and structural integrity. These efforts will be essential for meeting the demands of next-generation CN and VCN instrumentation and for complementing ongoing advancements in high-flux neutron sources.

\begin{acknowledgments}
This work is based on experiments performed at the Swiss spallation neutron source SINQ, Paul Scherrer Institute, Villigen, Switzerland.
We acknowledge financial support by the ILL and Vienna Doctoral School in Physics (VDSP), in particular a VDSP mobility fellowship provided to EH.  
We wish to acknowledge financial support of the Ministry of Education, Culture, Sports, Science and Technology of Japan (No. 15H03576). 
This research was funded in part by the Austrian Research
Promotion Agency (FFG), Quantum-Austria NextPi, grant number FO999896034 and the European Union: "NextGenerationEU". 
This research was funded in part by the Austrian Science Fund (FWF) [P-35597-N]. 
For the purpose of open access, the author has applied a CC BY public copyright licence to any Author Accepted Manuscript version arising from this submission.
\end{acknowledgments}

\section*{Data Availability Statement}

Raw data were generated at the Paul Scherrer Institute large scale facility. Derived data supporting the findings of this study are available from the corresponding author upon reasonable request.

\nocite{*}
\bibliography{EH_HBP_paper}

\providecommand{\noopsort}[1]{}\providecommand{\singleletter}[1]{#1}%
\begin{thebibliography}{21}%
\makeatletter
\providecommand \@ifxundefined [1]{%
 \@ifx{#1\undefined}
}%
\providecommand \@ifnum [1]{%
 \ifnum #1\expandafter \@firstoftwo
 \else \expandafter \@secondoftwo
 \fi
}%
\providecommand \@ifx [1]{%
 \ifx #1\expandafter \@firstoftwo
 \else \expandafter \@secondoftwo
 \fi
}%
\providecommand \natexlab [1]{#1}%
\providecommand \enquote  [1]{``#1''}%
\providecommand \bibnamefont  [1]{#1}%
\providecommand \bibfnamefont [1]{#1}%
\providecommand \citenamefont [1]{#1}%
\providecommand \href@noop [0]{\@secondoftwo}%
\providecommand \href [0]{\begingroup \@sanitize@url \@href}%
\providecommand \@href[1]{\@@startlink{#1}\@@href}%
\providecommand \@@href[1]{\endgroup#1\@@endlink}%
\providecommand \@sanitize@url [0]{\catcode `\\12\catcode `\$12\catcode `\&12\catcode `\#12\catcode `\^12\catcode `\_12\catcode `\%12\relax}%
\providecommand \@@startlink[1]{}%
\providecommand \@@endlink[0]{}%
\providecommand \url  [0]{\begingroup\@sanitize@url \@url }%
\providecommand \@url [1]{\endgroup\@href {#1}{\urlprefix }}%
\providecommand \urlprefix  [0]{URL }%
\providecommand \Eprint [0]{\href }%
\providecommand \doibase [0]{https://doi.org/}%
\providecommand \selectlanguage [0]{\@gobble}%
\providecommand \bibinfo  [0]{\@secondoftwo}%
\providecommand \bibfield  [0]{\@secondoftwo}%
\providecommand \translation [1]{[#1]}%
\providecommand \BibitemOpen [0]{}%
\providecommand \bibitemStop [0]{}%
\providecommand \bibitemNoStop [0]{.\EOS\space}%
\providecommand \EOS [0]{\spacefactor3000\relax}%
\providecommand \BibitemShut  [1]{\csname bibitem#1\endcsname}%
\let\auto@bib@innerbib\@empty
\bibitem [{\citenamefont {Rauch}\ and\ \citenamefont {Werner}(2015)}]{2015_RauchWerner_Book}%
  \BibitemOpen
  \bibfield  {author} {\bibinfo {author} {\bibfnamefont {H.}~\bibnamefont {Rauch}}\ and\ \bibinfo {author} {\bibfnamefont {S.~A.}\ \bibnamefont {Werner}},\ }\href@noop {} {\emph {\bibinfo {title} {Neutron Interferometry: Lessons in Experimental Quantum Mechanics, Wave-Particle Duality, and Entanglement}}}\ (\bibinfo  {publisher} {Oxford University Press},\ \bibinfo {address} {Oxford, UK},\ \bibinfo {year} {2015})\BibitemShut {NoStop}%
\bibitem [{\citenamefont {Haun}\ \emph {et~al.}(2020)\citenamefont {Haun}, \citenamefont {Wietfeldt}, \citenamefont {Arif}, \citenamefont {Huber}, \citenamefont {Black}, \citenamefont {Heacock}, \citenamefont {Pushin},\ and\ \citenamefont {Shahi}}]{2020_Haun_PRL}%
  \BibitemOpen
  \bibfield  {author} {\bibinfo {author} {\bibfnamefont {R.}~\bibnamefont {Haun}}, \bibinfo {author} {\bibfnamefont {F.~E.}\ \bibnamefont {Wietfeldt}}, \bibinfo {author} {\bibfnamefont {M.}~\bibnamefont {Arif}}, \bibinfo {author} {\bibfnamefont {M.~G.}\ \bibnamefont {Huber}}, \bibinfo {author} {\bibfnamefont {T.~C.}\ \bibnamefont {Black}}, \bibinfo {author} {\bibfnamefont {B.}~\bibnamefont {Heacock}}, \bibinfo {author} {\bibfnamefont {D.~A.}\ \bibnamefont {Pushin}},\ and\ \bibinfo {author} {\bibfnamefont {C.~B.}\ \bibnamefont {Shahi}},\ }\bibfield  {title} {\enquote {\bibinfo {title} {Precision measurement of the neutron scattering length of $^{4}\mathrm{He}$ using neutron interferometry},}\ }\href {https://doi.org/10.1103/PhysRevLett.124.012501} {\bibfield  {journal} {\bibinfo  {journal} {Phys. Rev. Lett.}\ }\textbf {\bibinfo {volume} {124}},\ \bibinfo {pages} {012501} (\bibinfo {year} {2020})}\BibitemShut {NoStop}%
\bibitem [{\citenamefont {Danner}\ \emph {et~al.}(2024)\citenamefont {Danner}, \citenamefont {Geerits}, \citenamefont {Lemmel}, \citenamefont {Wagner}, \citenamefont {Sponar},\ and\ \citenamefont {Hasegawa}}]{2024_Danner_Nat}%
  \BibitemOpen
  \bibfield  {author} {\bibinfo {author} {\bibfnamefont {A.}~\bibnamefont {Danner}}, \bibinfo {author} {\bibfnamefont {N.}~\bibnamefont {Geerits}}, \bibinfo {author} {\bibfnamefont {H.}~\bibnamefont {Lemmel}}, \bibinfo {author} {\bibfnamefont {R.}~\bibnamefont {Wagner}}, \bibinfo {author} {\bibfnamefont {S.}~\bibnamefont {Sponar}},\ and\ \bibinfo {author} {\bibfnamefont {Y.}~\bibnamefont {Hasegawa}},\ }\bibfield  {title} {\enquote {\bibinfo {title} {Three-path quantum cheshire cat observed in neutron interferometry},}\ }\href {https://doi.org/10.1038/s42005-023-01494-5} {\bibfield  {journal} {\bibinfo  {journal} {Communications Physics}\ }\textbf {\bibinfo {volume} {7}},\ \bibinfo {pages} {14} (\bibinfo {year} {2024})}\BibitemShut {NoStop}%
\bibitem [{\citenamefont {Eder}\ \emph {et~al.}(1989)\citenamefont {Eder}, \citenamefont {Gruber}, \citenamefont {Zeilinger}, \citenamefont {G{\"a}hler}, \citenamefont {Mampe},\ and\ \citenamefont {Drexel}}]{1989_Eder_NIMA}%
  \BibitemOpen
  \bibfield  {author} {\bibinfo {author} {\bibfnamefont {K.}~\bibnamefont {Eder}}, \bibinfo {author} {\bibfnamefont {M.}~\bibnamefont {Gruber}}, \bibinfo {author} {\bibfnamefont {A.}~\bibnamefont {Zeilinger}}, \bibinfo {author} {\bibfnamefont {R.}~\bibnamefont {G{\"a}hler}}, \bibinfo {author} {\bibfnamefont {W.}~\bibnamefont {Mampe}},\ and\ \bibinfo {author} {\bibfnamefont {W.}~\bibnamefont {Drexel}},\ }\bibfield  {title} {\enquote {\bibinfo {title} {The new very-cold-neutron optics facility at {ILL}},}\ }\href {https://doi.org/10.1016/0168-9002(89)90273-8} {\bibfield  {journal} {\bibinfo  {journal} {Nucl. Instrum. Methods Phys. Res., Sect. A}\ }\textbf {\bibinfo {volume} {284}},\ \bibinfo {pages} {171--175} (\bibinfo {year} {1989})}\BibitemShut {NoStop}%
\bibitem [{\citenamefont {Mezei}(2022)}]{2010_Mezei_JNR}%
  \BibitemOpen
  \bibfield  {author} {\bibinfo {author} {\bibfnamefont {F.}~\bibnamefont {Mezei}},\ }\bibfield  {title} {\enquote {\bibinfo {title} {Very cold neutrons in condensed matter research},}\ }\href {https://journals.sagepub.com/doi/full/10.3233/JNR-220012} {\bibfield  {journal} {\bibinfo  {journal} {Journal of Neutron Research}\ }\textbf {\bibinfo {volume} {24}},\ \bibinfo {pages} {205--210} (\bibinfo {year} {2022})}\BibitemShut {NoStop}%
\bibitem [{\citenamefont {Santoro}\ \emph {et~al.}(2024)\citenamefont {Santoro}, \citenamefont {Andersen}, \citenamefont {Bentley}, \citenamefont {Bernasconi}, \citenamefont {Bertelsen}, \citenamefont {Be\ss{}ler}, \citenamefont {Bianchi}, \citenamefont {Brys}, \citenamefont {Campi}, \citenamefont {Chambon}, \citenamefont {Czamler}, \citenamefont {Julio}, \citenamefont {Dian}, \citenamefont {Dunne}, \citenamefont {Ferreira}, \citenamefont {Fierlinger}, \citenamefont {Friman-Gayer}, \citenamefont {Folsom}, \citenamefont {Gaye}, \citenamefont {Gorini}, \citenamefont {Happe}, \citenamefont {Holl}, \citenamefont {Kamyshkov}, \citenamefont {Kittelmann}, \citenamefont {Klinkby}, \citenamefont {Kolevatov}, \citenamefont {Laporte}, \citenamefont {Lauritzen}, \citenamefont {Damian}, \citenamefont {Meirose}, \citenamefont {Mezei}, \citenamefont {Milstead}, \citenamefont {Muhrer}, \citenamefont {Neshvizhevsky}, \citenamefont {Rataj}, \citenamefont {Rizzi}, \citenamefont {Rosta}, \citenamefont {Samothrakitis},
  \citenamefont {Schober}, \citenamefont {Selknaes}, \citenamefont {Silverstein}, \citenamefont {Strobl}, \citenamefont {Strothmann}, \citenamefont {Takibayev}, \citenamefont {Wagner}, \citenamefont {Willendrup}, \citenamefont {Xu}, \citenamefont {Yiu}, \citenamefont {Zanini},\ and\ \citenamefont {and}}]{2024_Santoro_NSE}%
  \BibitemOpen
  \bibfield  {author} {\bibinfo {author} {\bibfnamefont {V.}~\bibnamefont {Santoro}}, \bibinfo {author} {\bibfnamefont {K.~H.}\ \bibnamefont {Andersen}}, \bibinfo {author} {\bibfnamefont {P.}~\bibnamefont {Bentley}}, \bibinfo {author} {\bibfnamefont {M.}~\bibnamefont {Bernasconi}}, \bibinfo {author} {\bibfnamefont {M.}~\bibnamefont {Bertelsen}}, \bibinfo {author} {\bibfnamefont {Y.}~\bibnamefont {Be\ss{}ler}}, \bibinfo {author} {\bibfnamefont {A.}~\bibnamefont {Bianchi}}, \bibinfo {author} {\bibfnamefont {T.}~\bibnamefont {Brys}}, \bibinfo {author} {\bibfnamefont {D.}~\bibnamefont {Campi}}, \bibinfo {author} {\bibfnamefont {A.}~\bibnamefont {Chambon}}, \bibinfo {author} {\bibfnamefont {V.}~\bibnamefont {Czamler}}, \bibinfo {author} {\bibfnamefont {D.~D.~D.}\ \bibnamefont {Julio}}, \bibinfo {author} {\bibfnamefont {E.}~\bibnamefont {Dian}}, \bibinfo {author} {\bibfnamefont {K.}~\bibnamefont {Dunne}}, \bibinfo {author} {\bibfnamefont {M.~J.}\ \bibnamefont {Ferreira}}, \bibinfo {author} {\bibfnamefont
  {P.}~\bibnamefont {Fierlinger}}, \bibinfo {author} {\bibfnamefont {U.}~\bibnamefont {Friman-Gayer}}, \bibinfo {author} {\bibfnamefont {B.~T.}\ \bibnamefont {Folsom}}, \bibinfo {author} {\bibfnamefont {A.}~\bibnamefont {Gaye}}, \bibinfo {author} {\bibfnamefont {G.}~\bibnamefont {Gorini}}, \bibinfo {author} {\bibfnamefont {C.}~\bibnamefont {Happe}}, \bibinfo {author} {\bibfnamefont {M.}~\bibnamefont {Holl}}, \bibinfo {author} {\bibfnamefont {Y.}~\bibnamefont {Kamyshkov}}, \bibinfo {author} {\bibfnamefont {T.}~\bibnamefont {Kittelmann}}, \bibinfo {author} {\bibfnamefont {E.~B.}\ \bibnamefont {Klinkby}}, \bibinfo {author} {\bibfnamefont {R.}~\bibnamefont {Kolevatov}}, \bibinfo {author} {\bibfnamefont {S.~I.}\ \bibnamefont {Laporte}}, \bibinfo {author} {\bibfnamefont {B.}~\bibnamefont {Lauritzen}}, \bibinfo {author} {\bibfnamefont {J.~I.~M.}\ \bibnamefont {Damian}}, \bibinfo {author} {\bibfnamefont {B.}~\bibnamefont {Meirose}}, \bibinfo {author} {\bibfnamefont {F.}~\bibnamefont {Mezei}}, \bibinfo {author}
  {\bibfnamefont {D.}~\bibnamefont {Milstead}}, \bibinfo {author} {\bibfnamefont {G.}~\bibnamefont {Muhrer}}, \bibinfo {author} {\bibfnamefont {V.}~\bibnamefont {Neshvizhevsky}}, \bibinfo {author} {\bibfnamefont {B.}~\bibnamefont {Rataj}}, \bibinfo {author} {\bibfnamefont {N.}~\bibnamefont {Rizzi}}, \bibinfo {author} {\bibfnamefont {L.}~\bibnamefont {Rosta}}, \bibinfo {author} {\bibfnamefont {S.}~\bibnamefont {Samothrakitis}}, \bibinfo {author} {\bibfnamefont {H.}~\bibnamefont {Schober}}, \bibinfo {author} {\bibfnamefont {J.~R.}\ \bibnamefont {Selknaes}}, \bibinfo {author} {\bibfnamefont {S.}~\bibnamefont {Silverstein}}, \bibinfo {author} {\bibfnamefont {M.}~\bibnamefont {Strobl}}, \bibinfo {author} {\bibfnamefont {M.}~\bibnamefont {Strothmann}}, \bibinfo {author} {\bibfnamefont {A.}~\bibnamefont {Takibayev}}, \bibinfo {author} {\bibfnamefont {R.}~\bibnamefont {Wagner}}, \bibinfo {author} {\bibfnamefont {P.}~\bibnamefont {Willendrup}}, \bibinfo {author} {\bibfnamefont {S.}~\bibnamefont {Xu}}, \bibinfo
  {author} {\bibfnamefont {S.~C.}\ \bibnamefont {Yiu}}, \bibinfo {author} {\bibfnamefont {L.}~\bibnamefont {Zanini}},\ and\ \bibinfo {author} {\bibfnamefont {O.~Z.}\ \bibnamefont {and}},\ }\bibfield  {title} {\enquote {\bibinfo {title} {The {HighNESS} project at the european spallation source: Current status and future perspectives},}\ }\href {https://doi.org/10.1080/00295639.2023.2204184} {\bibfield  {journal} {\bibinfo  {journal} {Nuclear Science and Engineering}\ }\textbf {\bibinfo {volume} {198}},\ \bibinfo {pages} {31--63} (\bibinfo {year} {2024})}\BibitemShut {NoStop}%
\bibitem [{\citenamefont {Rupp}\ \emph {et~al.}(1990)\citenamefont {Rupp}, \citenamefont {Hehmann}, \citenamefont {Matull},\ and\ \citenamefont {Ibel}}]{1990_Rupp_PRL}%
  \BibitemOpen
  \bibfield  {author} {\bibinfo {author} {\bibfnamefont {R.~A.}\ \bibnamefont {Rupp}}, \bibinfo {author} {\bibfnamefont {J.}~\bibnamefont {Hehmann}}, \bibinfo {author} {\bibfnamefont {R.}~\bibnamefont {Matull}},\ and\ \bibinfo {author} {\bibfnamefont {K.}~\bibnamefont {Ibel}},\ }\bibfield  {title} {\enquote {\bibinfo {title} {Neutron diffraction from photoinduced gratings in a {PMMA} matrix},}\ }\href {https://doi.org/10.1103/PhysRevLett.64.301} {\bibfield  {journal} {\bibinfo  {journal} {Phys. Rev. Lett.}\ }\textbf {\bibinfo {volume} {64}},\ \bibinfo {pages} {301--302} (\bibinfo {year} {1990})}\BibitemShut {NoStop}%
\bibitem [{\citenamefont {Fally}\ \emph {et~al.}(2006)\citenamefont {Fally}, \citenamefont {Drevensek-Olenik}, \citenamefont {Ellabban}, \citenamefont {Pranzas},\ and\ \citenamefont {Vollbrandt}}]{2006_Fally_PRL}%
  \BibitemOpen
  \bibfield  {author} {\bibinfo {author} {\bibfnamefont {M.}~\bibnamefont {Fally}}, \bibinfo {author} {\bibfnamefont {I.}~\bibnamefont {Drevensek-Olenik}}, \bibinfo {author} {\bibfnamefont {M.}~\bibnamefont {Ellabban}}, \bibinfo {author} {\bibfnamefont {K.}~\bibnamefont {Pranzas}},\ and\ \bibinfo {author} {\bibfnamefont {J.}~\bibnamefont {Vollbrandt}},\ }\bibfield  {title} {\enquote {\bibinfo {title} {Colossal light-induced refractive-index modulation for neutrons in holographic polymer-dispersed liquid crystals},}\ }\href {https://doi.org/10.1103/PHYSREVLETT.97.167803} {\bibfield  {journal} {\bibinfo  {journal} {Phys. Rev. Lett.}\ }\textbf {\bibinfo {volume} {97}},\ \bibinfo {pages} {167803} (\bibinfo {year} {2006})}\BibitemShut {NoStop}%
\bibitem [{\citenamefont {Tomita}\ \emph {et~al.}(2016{\natexlab{a}})\citenamefont {Tomita}, \citenamefont {Hata}, \citenamefont {Momose}, \citenamefont {Takayama}, \citenamefont {Liu}, \citenamefont {Chikama}, \citenamefont {Klepp}, \citenamefont {Pruner},\ and\ \citenamefont {and}}]{2016_Tomita_JMO}%
  \BibitemOpen
  \bibfield  {author} {\bibinfo {author} {\bibfnamefont {Y.}~\bibnamefont {Tomita}}, \bibinfo {author} {\bibfnamefont {E.}~\bibnamefont {Hata}}, \bibinfo {author} {\bibfnamefont {K.}~\bibnamefont {Momose}}, \bibinfo {author} {\bibfnamefont {S.}~\bibnamefont {Takayama}}, \bibinfo {author} {\bibfnamefont {X.}~\bibnamefont {Liu}}, \bibinfo {author} {\bibfnamefont {K.}~\bibnamefont {Chikama}}, \bibinfo {author} {\bibfnamefont {J.}~\bibnamefont {Klepp}}, \bibinfo {author} {\bibfnamefont {C.}~\bibnamefont {Pruner}},\ and\ \bibinfo {author} {\bibfnamefont {M.~F.}\ \bibnamefont {and}},\ }\bibfield  {title} {\enquote {\bibinfo {title} {Photopolymerizable nanocomposite photonic materials and their holographic applications in light and neutron optics},}\ }\href {https://doi.org/10.1080/09500340.2016.1143534} {\bibfield  {journal} {\bibinfo  {journal} {J. Mod. Opt.}\ }\textbf {\bibinfo {volume} {63}},\ \bibinfo {pages} {S1--S31} (\bibinfo {year} {2016}{\natexlab{a}})}\BibitemShut {NoStop}%
\bibitem [{\citenamefont {Fally}\ \emph {et~al.}(2010)\citenamefont {Fally}, \citenamefont {Klepp}, \citenamefont {Tomita}, \citenamefont {Nakamura}, \citenamefont {Pruner}, \citenamefont {Ellabban}, \citenamefont {Rupp}, \citenamefont {Bichler}, \citenamefont {Drevensek-Olenik}, \citenamefont {Kohlbrecher}, \citenamefont {Eckerlebe}, \citenamefont {Lemmel},\ and\ \citenamefont {Rauch}}]{2010_Fally_PRL_NPCs}%
  \BibitemOpen
  \bibfield  {author} {\bibinfo {author} {\bibfnamefont {M.}~\bibnamefont {Fally}}, \bibinfo {author} {\bibfnamefont {J.}~\bibnamefont {Klepp}}, \bibinfo {author} {\bibfnamefont {Y.}~\bibnamefont {Tomita}}, \bibinfo {author} {\bibfnamefont {T.}~\bibnamefont {Nakamura}}, \bibinfo {author} {\bibfnamefont {C.}~\bibnamefont {Pruner}}, \bibinfo {author} {\bibfnamefont {M.~A.}\ \bibnamefont {Ellabban}}, \bibinfo {author} {\bibfnamefont {R.~A.}\ \bibnamefont {Rupp}}, \bibinfo {author} {\bibfnamefont {M.}~\bibnamefont {Bichler}}, \bibinfo {author} {\bibfnamefont {I.}~\bibnamefont {Drevensek-Olenik}}, \bibinfo {author} {\bibfnamefont {J.}~\bibnamefont {Kohlbrecher}}, \bibinfo {author} {\bibfnamefont {H.}~\bibnamefont {Eckerlebe}}, \bibinfo {author} {\bibfnamefont {H.}~\bibnamefont {Lemmel}},\ and\ \bibinfo {author} {\bibfnamefont {H.}~\bibnamefont {Rauch}},\ }\bibfield  {title} {\enquote {\bibinfo {title} {Neutron optical beam splitter from holographically structured nanoparticle-polymer composites},}\ }\href
  {https://doi.org/10.1103/PhysRevLett.105.123904} {\bibfield  {journal} {\bibinfo  {journal} {Phys. Rev. Lett.}\ }\textbf {\bibinfo {volume} {105}},\ \bibinfo {pages} {123904} (\bibinfo {year} {2010})}\BibitemShut {NoStop}%
\bibitem [{\citenamefont {Tomita}\ \emph {et~al.}(2020{\natexlab{a}})\citenamefont {Tomita}, \citenamefont {Kageyama}, \citenamefont {Iso}, \citenamefont {Umemoto}, \citenamefont {Kume}, \citenamefont {Liu}, \citenamefont {Pruner}, \citenamefont {Jenke}, \citenamefont {Roccia}, \citenamefont {Geltenbort}, \citenamefont {Fally},\ and\ \citenamefont {Klepp}}]{2020_Tomita_PRA}%
  \BibitemOpen
  \bibfield  {author} {\bibinfo {author} {\bibfnamefont {Y.}~\bibnamefont {Tomita}}, \bibinfo {author} {\bibfnamefont {A.}~\bibnamefont {Kageyama}}, \bibinfo {author} {\bibfnamefont {Y.}~\bibnamefont {Iso}}, \bibinfo {author} {\bibfnamefont {K.}~\bibnamefont {Umemoto}}, \bibinfo {author} {\bibfnamefont {A.}~\bibnamefont {Kume}}, \bibinfo {author} {\bibfnamefont {M.}~\bibnamefont {Liu}}, \bibinfo {author} {\bibfnamefont {C.}~\bibnamefont {Pruner}}, \bibinfo {author} {\bibfnamefont {T.}~\bibnamefont {Jenke}}, \bibinfo {author} {\bibfnamefont {S.}~\bibnamefont {Roccia}}, \bibinfo {author} {\bibfnamefont {P.}~\bibnamefont {Geltenbort}}, \bibinfo {author} {\bibfnamefont {M.}~\bibnamefont {Fally}},\ and\ \bibinfo {author} {\bibfnamefont {J.}~\bibnamefont {Klepp}},\ }\bibfield  {title} {\enquote {\bibinfo {title} {Fabrication of nanodiamond-dispersed composite holographic gratings and their light and slow-neutron diffraction properties},}\ }\href {https://doi.org/10.1103/PhysRevApplied.14.044056} {\bibfield
  {journal} {\bibinfo  {journal} {Phys. Rev. Applied}\ }\textbf {\bibinfo {volume} {14}},\ \bibinfo {pages} {044056} (\bibinfo {year} {2020}{\natexlab{a}})}\BibitemShut {NoStop}%
\bibitem [{\citenamefont {Hadden}\ \emph {et~al.}(2022)\citenamefont {Hadden}, \citenamefont {Iso}, \citenamefont {Kume}, \citenamefont {Umemoto}, \citenamefont {Jenke}, \citenamefont {Fally}, \citenamefont {Klepp},\ and\ \citenamefont {Tomita}}]{2022_Hadden_SPIE}%
  \BibitemOpen
  \bibfield  {author} {\bibinfo {author} {\bibfnamefont {E.}~\bibnamefont {Hadden}}, \bibinfo {author} {\bibfnamefont {Y.}~\bibnamefont {Iso}}, \bibinfo {author} {\bibfnamefont {A.}~\bibnamefont {Kume}}, \bibinfo {author} {\bibfnamefont {K.}~\bibnamefont {Umemoto}}, \bibinfo {author} {\bibfnamefont {T.}~\bibnamefont {Jenke}}, \bibinfo {author} {\bibfnamefont {M.}~\bibnamefont {Fally}}, \bibinfo {author} {\bibfnamefont {J.}~\bibnamefont {Klepp}},\ and\ \bibinfo {author} {\bibfnamefont {Y.}~\bibnamefont {Tomita}},\ }\bibfield  {title} {\enquote {\bibinfo {title} {Nanodiamond-based nanoparticle-polymer composite gratings with extremely large neutron refractive index modulation},}\ }in\ \href {https://doi.org/10.1117/12.2623661} {\emph {\bibinfo {booktitle} {Photosensitive Materials and Their Applications II}}},\ Vol.\ \bibinfo {volume} {12151}\ (\bibinfo {organization} {SPIE},\ \bibinfo {year} {2022})\ pp.\ \bibinfo {pages} {70--76}\BibitemShut {NoStop}%
\bibitem [{\citenamefont {Hadden}\ \emph {et~al.}(2024)\citenamefont {Hadden}, \citenamefont {Fally}, \citenamefont {Iso}, \citenamefont {Jenke}, \citenamefont {Klepp}, \citenamefont {Kume}, \citenamefont {Umemoto},\ and\ \citenamefont {Tomita}}]{2024_Hadden_APL}%
  \BibitemOpen
  \bibfield  {author} {\bibinfo {author} {\bibfnamefont {E.}~\bibnamefont {Hadden}}, \bibinfo {author} {\bibfnamefont {M.}~\bibnamefont {Fally}}, \bibinfo {author} {\bibfnamefont {Y.}~\bibnamefont {Iso}}, \bibinfo {author} {\bibfnamefont {T.}~\bibnamefont {Jenke}}, \bibinfo {author} {\bibfnamefont {J.}~\bibnamefont {Klepp}}, \bibinfo {author} {\bibfnamefont {A.}~\bibnamefont {Kume}}, \bibinfo {author} {\bibfnamefont {K.}~\bibnamefont {Umemoto}},\ and\ \bibinfo {author} {\bibfnamefont {Y.}~\bibnamefont {Tomita}},\ }\bibfield  {title} {\enquote {\bibinfo {title} {{Holographic nanodiamond–polymer composite grating with unprecedented slow-neutron refractive index modulation amplitude}},}\ }\href {https://doi.org/10.1063/5.0186753} {\bibfield  {journal} {\bibinfo  {journal} {Appl. Phys. Lett.}\ }\textbf {\bibinfo {volume} {124}},\ \bibinfo {pages} {071901} (\bibinfo {year} {2024})}\BibitemShut {NoStop}%
\bibitem [{\citenamefont {Tomita}\ \emph {et~al.}(2016{\natexlab{b}})\citenamefont {Tomita}, \citenamefont {Urano}, \citenamefont {Fukamizu}, \citenamefont {Kametani}, \citenamefont {Nishimura},\ and\ \citenamefont {Odoi}}]{2016_Tomita_OL_HBPs}%
  \BibitemOpen
  \bibfield  {author} {\bibinfo {author} {\bibfnamefont {Y.}~\bibnamefont {Tomita}}, \bibinfo {author} {\bibfnamefont {H.}~\bibnamefont {Urano}}, \bibinfo {author} {\bibfnamefont {T.}~\bibnamefont {Fukamizu}}, \bibinfo {author} {\bibfnamefont {Y.}~\bibnamefont {Kametani}}, \bibinfo {author} {\bibfnamefont {N.}~\bibnamefont {Nishimura}},\ and\ \bibinfo {author} {\bibfnamefont {K.}~\bibnamefont {Odoi}},\ }\bibfield  {title} {\enquote {\bibinfo {title} {Nanoparticle-polymer composite volume holographic gratings dispersed with ultrahigh-refractive-index hyperbranched polymer as organic nanoparticles},}\ }\href {https://doi.org/10.1364/OL.41.001281} {\bibfield  {journal} {\bibinfo  {journal} {Opt. Lett.}\ }\textbf {\bibinfo {volume} {41}},\ \bibinfo {pages} {1281--1284} (\bibinfo {year} {2016}{\natexlab{b}})}\BibitemShut {NoStop}%
\bibitem [{\citenamefont {Tomita}\ \emph {et~al.}(2020{\natexlab{b}})\citenamefont {Tomita}, \citenamefont {Aoi}, \citenamefont {Hasegawa}, \citenamefont {Xia}, \citenamefont {Wang},\ and\ \citenamefont {Oshima}}]{2020_Tomita_OE_HBP}%
  \BibitemOpen
  \bibfield  {author} {\bibinfo {author} {\bibfnamefont {Y.}~\bibnamefont {Tomita}}, \bibinfo {author} {\bibfnamefont {T.}~\bibnamefont {Aoi}}, \bibinfo {author} {\bibfnamefont {S.}~\bibnamefont {Hasegawa}}, \bibinfo {author} {\bibfnamefont {F.}~\bibnamefont {Xia}}, \bibinfo {author} {\bibfnamefont {Y.}~\bibnamefont {Wang}},\ and\ \bibinfo {author} {\bibfnamefont {J.}~\bibnamefont {Oshima}},\ }\bibfield  {title} {\enquote {\bibinfo {title} {Very high contrast volume holographic gratings recorded in photopolymerizable nanocomposite materials},}\ }\href {https://doi.org/10.1364/OE.400092} {\bibfield  {journal} {\bibinfo  {journal} {Opt. Express}\ }\textbf {\bibinfo {volume} {28}},\ \bibinfo {pages} {28366--28382} (\bibinfo {year} {2020}{\natexlab{b}})}\BibitemShut {NoStop}%
\bibitem [{\citenamefont {Narita}\ \emph {et~al.}(2021)\citenamefont {Narita}, \citenamefont {Oshima}, \citenamefont {Iso}, \citenamefont {Hasegawa},\ and\ \citenamefont {Tomita}}]{2021_Narita_OME}%
  \BibitemOpen
  \bibfield  {author} {\bibinfo {author} {\bibfnamefont {A.}~\bibnamefont {Narita}}, \bibinfo {author} {\bibfnamefont {J.}~\bibnamefont {Oshima}}, \bibinfo {author} {\bibfnamefont {Y.}~\bibnamefont {Iso}}, \bibinfo {author} {\bibfnamefont {S.}~\bibnamefont {Hasegawa}},\ and\ \bibinfo {author} {\bibfnamefont {Y.}~\bibnamefont {Tomita}},\ }\bibfield  {title} {\enquote {\bibinfo {title} {Red-sensitive organic nanoparticle-polymer composite materials for volume holographic gratings with large refractive index modulation amplitudes},}\ }\href {https://doi.org/10.1364/OME.415422} {\bibfield  {journal} {\bibinfo  {journal} {Opt. Mater. Express}\ }\textbf {\bibinfo {volume} {11}},\ \bibinfo {pages} {614--628} (\bibinfo {year} {2021})}\BibitemShut {NoStop}%
\bibitem [{\citenamefont {Uchida}(1973)}]{1973_Uchida_JOSA}%
  \BibitemOpen
  \bibfield  {author} {\bibinfo {author} {\bibfnamefont {N.}~\bibnamefont {Uchida}},\ }\bibfield  {title} {\enquote {\bibinfo {title} {Calculation of diffraction efficiency in hologram gratings attenuated along the direction perpendicular to the grating vector},}\ }\href {https://doi.org/10.1364/JOSA.63.000280} {\bibfield  {journal} {\bibinfo  {journal} {J. Opt. Soc. Am.}\ }\textbf {\bibinfo {volume} {63}},\ \bibinfo {pages} {280--287} (\bibinfo {year} {1973})}\BibitemShut {NoStop}%
\bibitem [{\citenamefont {Sheridan}(1992)}]{1992_Sheridan_JMO}%
  \BibitemOpen
  \bibfield  {author} {\bibinfo {author} {\bibfnamefont {J.}~\bibnamefont {Sheridan}},\ }\bibfield  {title} {\enquote {\bibinfo {title} {A comparison of diffraction theories for {off-Bragg} replay},}\ }\href {https://doi.org/10.1080/713823578} {\bibfield  {journal} {\bibinfo  {journal} {Journal of Modern Optics}\ }\textbf {\bibinfo {volume} {39}},\ \bibinfo {pages} {1709--1718} (\bibinfo {year} {1992})}\BibitemShut {NoStop}%
\bibitem [{\citenamefont {Suzuki}\ and\ \citenamefont {Tomita}(2007)}]{2007_Suzuki}%
  \BibitemOpen
  \bibfield  {author} {\bibinfo {author} {\bibfnamefont {N.}~\bibnamefont {Suzuki}}\ and\ \bibinfo {author} {\bibfnamefont {Y.}~\bibnamefont {Tomita}},\ }\bibfield  {title} {\enquote {\bibinfo {title} {Holographic scattering in {SiO$_2$} nanoparticle-dispersed photopolymer films},}\ }\href {https://doi.org/10.1364/AO.46.006809} {\bibfield  {journal} {\bibinfo  {journal} {Appl. Opt.}\ }\textbf {\bibinfo {volume} {46}},\ \bibinfo {pages} {6809--6814} (\bibinfo {year} {2007})}\BibitemShut {NoStop}%
\bibitem [{\citenamefont {Hadden}(2024)}]{2024_Hadden_thesis}%
  \BibitemOpen
  \bibfield  {author} {\bibinfo {author} {\bibfnamefont {E.}~\bibnamefont {Hadden}},\ }\emph {\bibinfo {title} {Polymer based photonic materials for cold neutron optics}},\ \href {https://doi.org/10.25365/thesis.77766} {\bibinfo {type} {Dissertation}},\ \bibinfo  {school} {Universität Wien}, \bibinfo {address} {Wien} (\bibinfo {year} {2024})\BibitemShut {NoStop}%
\bibitem [{\citenamefont {Klepp}\ \emph {et~al.}(2016)\citenamefont {Klepp}, \citenamefont {Pruner}, \citenamefont {Tomita}, \citenamefont {Geltenbort}, \citenamefont {Kohlbrecher},\ and\ \citenamefont {Fally}}]{2016_Klepp_JPCS}%
  \BibitemOpen
  \bibfield  {author} {\bibinfo {author} {\bibfnamefont {J.}~\bibnamefont {Klepp}}, \bibinfo {author} {\bibfnamefont {C.}~\bibnamefont {Pruner}}, \bibinfo {author} {\bibfnamefont {Y.}~\bibnamefont {Tomita}}, \bibinfo {author} {\bibfnamefont {P.}~\bibnamefont {Geltenbort}}, \bibinfo {author} {\bibfnamefont {J.}~\bibnamefont {Kohlbrecher}},\ and\ \bibinfo {author} {\bibfnamefont {M.}~\bibnamefont {Fally}},\ }\bibfield  {title} {\enquote {\bibinfo {title} {Advancing data analysis for reflectivity measurements of holographic nanocomposite gratings},}\ }\href {https://doi.org/10.1088/1742-6596/746/1/012022} {\bibfield  {journal} {\bibinfo  {journal} {Journal of Physics: Conference Series}\ }\textbf {\bibinfo {volume} {746}},\ \bibinfo {pages} {012022} (\bibinfo {year} {2016})}\BibitemShut {NoStop}%
\end{thebibliography}%

\end{document}